\documentclass[aip,jmp]{revtex4-1}
\usepackage{amssymb}
\usepackage{graphics}

\begin{document}

\title{Regularization by Test Function}
\author{Peter Morgan}
\email{peter.w.morgan@yale.edu}
\affiliation{Physics Department, Yale University, New Haven, CT 06520, USA.}
\homepage{http://pantheon.yale.edu/~PWM22}

\date{\today}
\begin{abstract}
Quantum fields are generally taken to be operator--valued distributions, linear functionals of test functions into an algebra of operators; here the effective dynamics of an interacting quantum field is taken to be nonlinearly modified by properties of test functions, in a way that preserves Poincar\'e invariance, microcausality, and the Fock--Hilbert space structure of the free field. The construction can be taken to be a physically comprehensible regularization because we can introduce a sequence that has a limit that is a conventional interacting quantum field, with the usual informal dependence of the effective dynamics on properties of the experimental apparatus made formally explicit as a dependence on the test functions that are used to model the experimental apparatus.
\end{abstract}
\maketitle

\newcommand\Half{{\frac{1}{2}}}
\newcommand\Intd{{\mathrm{d}}}
\newcommand\Remove[1]{{\raise 1.2ex\hbox{$\times$}\kern-0.8em \lower 0.35ex\hbox{$#1$}}}
\newcommand\SmallFrac[2]{{\scriptstyle\frac{\scriptstyle #1}{\scriptstyle#2}}}
\newcommand\Vacuum{{\left|0\right>}}
\newcommand\VEV[1]{{\left<0\right|#1\left|0\right>}}
\newcommand\eqdef{{\stackrel{\scriptstyle\mathrm{def}}{=}}}
\newcommand\rme{{\mathrm{e}}}
\newcommand\rmi{{\mathrm{i}}}

\section{Introduction}
We work with a quantized real Klein-Gordon free field $\hat\phi(x)$ as an operator--valued distribution,
$$\hat\phi:\mathcal{S}\rightarrow\mathcal{A};f\mapsto\hat\phi_f
       =\int \hat\phi(x)f(x)\mathrm{d}^4x,\qquad f\in\mathcal{S},$$
where the test function space $\mathcal{S}$ is the Schwartz space of complex functions that are smooth both in real space and in fourier space, with commutator
$[\hat\phi_f,\hat\phi_g]=(f^*,g)-(g^*,f)$, which is zero when $f$ and $g$ have space--like separated supports, where the 2--point Vacuum Expectation Value (VEV) is the positive semi--definite inner product
$$(f,g)=\VEV{\hat\phi_f^\dagger\hat\phi_g}
  =\!\!\int\!\!\tilde f^*(k)\tilde g(k)
          2\pi\delta(k^2-m^2)\theta(k_0)\frac{\mathrm{d}^4k}{(2\pi)^4}.
$$
The vacuum state, defined as a linear map on the algebra of observables generated by $\hat\phi_f$ and recursively by
$$\VEV{\cdot}:\mathcal{A}\rightarrow\mathbb{C};\VEV{\hat\phi_{f_1}\cdots\hat\phi_{f_{n+1}}}=
   \sum_i(f_i^*,f_{n+1})\VEV{\hat\phi_{f_1}\cdots\Remove{\hat\phi_{f_i}}\cdots\hat\phi_{f_n}},
$$
is enough to allow the Gelfand-Naimark-Segal (GNS) construction of the Fock--Hilbert space\cite[\S III.2]{Haag}.

The construction here introduces a nonlinear dependence on test functions, so that we will introduce a nonlinear operator--valued functional $\hat\xi:\mathcal{S}\rightarrow\mathcal{A}; f\mapsto\hat\xi_f$ that approaches the quantum field operators of perturbative quantum field theory when the test functions approach the plane waves that we use to model experimental apparatus in High Energy Physics (HEP), but may be significantly different when the test functions are close to point--like in real space.
Although we will allow
$\hat\xi_{[\lambda_1 f_1+\lambda_2 f_2]}\not=\lambda_1\hat\xi_{f_1}+\lambda_2\hat\xi_{ f_2}$, we will require that the vacuum state over the algebra of observables that is generated by $\hat\xi_f$ is a linear functional,
$\VEV{[\lambda_1\hat A_1+\lambda_2\hat A_2]}=\lambda_1\VEV{\hat A_1}+\lambda_2\VEV{\hat A_2}$, so that a Hilbert space can still be GNS-constructed and a probability interpretation is still possible.

The particular mathematics proposed here is open to extension and refinement, but nonlinear quantum fields, including a less physically justified class of nonlinear quantum fields that has previously been suggested~\cite{MorganHadamard}, admit a significant range of nontrivial models.

\section{A test function perturbed dynamics}
As an approach to regularization, we will effectively use length scales of an envelope of a test function as a proxy for rescaling, in contrast to the conventional use of the highest frequency scale as a proxy for rescaling.
We first construct what we will call \emph{contracted envelopes} $f_{\lambda,x}$ of a complex test function $f$, centered on the point $x$ and with a relative scale functional $\lambda:\mathcal{S}\rightarrow \mathbb{R}_+$, for which the most important requirement is that $\mathrm{Supp}[f_{\lambda,x}]\subseteq\mathrm{Supp}[f]$.
For example, using the support of a function explicitly,
$$f_{\lambda,x}(y)=
  \left\{\begin{array}{l l}
    |f(\lambda[f](y-x)+x)|^2 &
        \quad y\in\mathrm{Supp}[f]\cr
   {} & {}\cr
   0 & \quad y\not\in\mathrm{Supp}[f]
  \end{array}\right.
$$
or, for an example that avoids using the support of the function,
$$f_{\lambda,x}(y)= |f(y)|^2 \cdot |f(\lambda[f](y-x)+x)|^2,
$$
or, for an example of an intermediate but more elaborate construction,
$$f_{\lambda,x}(y)=
  \frac{\mathrm{tanh}\hspace{-0.6ex}\left(\strut\hspace{-0.4ex}C|f(y)|^2\right)}
        {\mathrm{tanh}\hspace{-0.6ex}\left(\strut\hspace{-0.4ex}C\right)}
         \cdot |f(\lambda[f](y-x)+x)|^2,\qquad\mbox{ for some } C\in\mathbb{R}_+.
$$
$\lambda:\mathcal{S}\rightarrow \mathbb{R}_+$ is a positive real--valued Poincar\'e invariant functional, for which the most straightforward example, without introducing additional space--time structure, is to take some positive function $\lambda[f]=P_1\left((f,f)\strut\right)$, $P_1:\mathbb{R}_+\rightarrow\mathbb{R}_+$, but arbitrary complexity cannot be ruled out \emph{a priori}.
$f_{\lambda,x}$ is mathematically awkward for general smooth functions of bounded support, but it is relatively straightforward for a one--parameter set of analytic Schwartz space functions that approach a massive plane wave ---which is an appropriate starting point for reconstructing perturbative quantum field theory in momentum space--- such as
$$F(\mathsf{k},\mu;y)=\exp{\left(-\mathrm{i}\mathsf{k}\cdot y
           +\mu^2\left(y\cdot y-2\frac{(\mathsf{k}\cdot y)^2}{m^2}\right)\right)},
$$
where $\mu>0$ and $\mathsf{k}^2=m^2$ for a 4--vector parameter $\mathsf{k}$, for which
$\mathrm{Supp}[F(\mathsf{k},\mu;\cdot)]$ is the whole of Minkowski space.
We choose $P_1$ so that as $\mu\rightarrow 0$, the contracted envelope $F_{\lambda,x}(\mathsf{k},\mu;y)$ of $F(\mathsf{k},\mu;y)$, which for such a simple test function just extracts a Gaussian, approaches a multiple of a Dirac $\delta$--function in four dimensions as the envelope of $F(\mathsf{k},\mu;\cdot)$ becomes larger.

For the test function $f$, we construct an operator $\hat\xi_f$ that satisfies microcausality as a nonlinear functional of $f$ that acts on the free field Fock space, following closely the usual construction of an interacting quantum field (see Appendix \ref{IZAppendix}),
$$\hat\xi_f=\mathsf{T}^\dagger\!\left[\mathrm{e}^{-\mathrm{i}\hat\mathcal{L}[f]}\right]
        \mathsf{T}\!\!\left[\hat\phi_f\mathrm{e}^{-\mathrm{i}\hat\mathcal{L}[f]}\right],$$
where $\hat\mathcal{L}[f]$ is a self--adjoint operator that includes only free field operators $\hat\phi(z)$ for $z\in\mathrm{Supp}[f]$,
$$\hat\mathcal{L}[f]=\int\!\! :\!P_2\Biggl((f_{\lambda,x},f_{\lambda,x}),\hat\phi^{\ }_{f_{\lambda,x}}
  \hat\phi_{f_{\lambda,x}}^\dagger\Biggr)\!:\mathrm{d}^4x,\qquad
     P_2:\mathbb{R}_+\times\mathbb{R}_+\rightarrow\mathbb{R}_+,$$
because $\hat\phi^{\ }_{f_{\lambda,x}}=\int\hat\phi(z)f_{\lambda,x}(z)\mathrm{d}^4z$ has nonzero components only when $z\in\mathrm{Supp}[f]$.
As in Appendix \ref{IZAppendix}, this constructs a complex of free quantum field operators, but in this case confined to the support of the test function $f$ instead of being less confined, to the past light--cone of $f$.
This is not a significant restriction insofar as the Reeh-Schlieder theorem\cite[\S II.5.3]{Haag} shows that vector states generated by the algebra of observables $\mathcal{A}(\mathrm{Supp}[f])$ acting on the vacuum vector are dense in the Fock--Hilbert space of the free field, so that even where the support of a test function is small we should be able to use a sufficiently elaborate system of functionals of the test function to approximate whatever vector state can be constructed using free quantum field operators throughout the backward light--cone of the test function.

Except that $\hat\phi_f$ is an unbounded operator, so that, for example, expansion of the time--ordered exponential is not necessarily well--defined (a more technical difficulty than is usually presented by quantum field theory), this construction is well--defined for any test function $f\in\mathcal{S}$, for renormalizable or non--renormalizable Lagrangian densities, and the contracted envelopes of the test functions clearly regularize the momentum space Feynman integrals that are obtained, by smearing the operator--valued distributions $\hat\phi(x)$ within the interaction Lagrangian density.
For renormalizable Lagrangian densities we might want to choose $P_2$ so that $\hat\xi_{F(\mathsf{k},\mu;\cdot)}$ is well--defined in the limit $\mu\rightarrow 0$, however from a perspective of effective quantum field theory and of less idealized models of experimental apparatus we should investigate the response of a given theory to using different Schwartz space test functions to model physical states.

\section{Variations}
Two kinds of generalization are immediately possible if we introduce a parameterized set of contraction scale functions, $\lambda(\alpha):\mathcal{S}\rightarrow\mathbb{R}_+$ and a measure $\mathrm{d}\mu(\alpha)$.
We may, for example, introduce
$$f_{\lambda,x}(y)= |f(y)|^2\int|f(\lambda(\alpha)[f](y-x)+x)|^2\mathrm{d}\mu(\alpha),
$$
which smears the contracted envelope by using different scale factors.
The operator $\hat\mathcal{L}[f]$ may also be generalized to be a functional of a smeared local operator
$$\hat\Phi_f=\int\hat\phi^{\ }_{f_{\lambda(\alpha),x}}
         \hat\phi_{f_{\lambda(\alpha),x}}^\dagger\mathrm{d}\mu(\alpha),
$$
instead of being a functional of $\hat\phi^{\ }_{f_{\lambda,x}}\hat\phi_{f_{\lambda,x}}^\dagger$, which will also reduce the coherence of the dependency of the dynamics on the test function $f$.
These two constructions can be combined and made more elaborate, but, as always, questions of tractability are important when assessing the usefulness of a theory, even while we note that the response of the dynamics of any given theory to different test function envelopes (and the systematic rules we use to assign combinations of test functions to experimental preparation and measurement devices) will be open to experimental test.

Because all uses of the operator--valued distribution $\hat\phi(x)$ are smeared by a Schwartz space test function, it is possible to use a nonlinear construction introduced by the author elsewhere\cite{MorganHadamard} (for which, for the simplest example, $[\hat\phi_f,\hat\phi_g]=(f^*,g)^n-(g^*,f)^n$ and $\VEV{\hat\phi_f^\dagger\hat\phi_g}=(f,g)^n$), although there is no immediate physical motivation for doing so.

We could equally well use $f_{\lambda,x}(y)=f(y)f(\lambda[f](y-x)+x)$, without taking the absolute value, which does not extract an envelope for the plane wave test function case $F(\mathsf{k},\mu;y)$.
With this construction $F_{\lambda,x}(\mathsf{k},\mu;y)$ has increasingly high frequency components as $\mu\rightarrow 0$, instead of being a simple Gaussian, so that we do not obtain a straightforward regularization of fourier space Feynman integrals.
Such a construction cannot be ruled out \emph{a priori}, indeed it makes a reasonable suggestion, that all properties of an experimental apparatus have a nonlinear effect on the dynamics, not just the envelopes of the test functions that are used in a model, but it introduces a relatively large step away from conventional HEP.

We can also introduce $U(1)$--contravariant test functions and connections that allow the construction of $U(1)$--invariant Maxwell--Dirac--type theories, however despite the requirement of $U(1)$--invariant VEVs there are even more possible ways to smear the operator--valued operators in the interaction Lagrangian than for the scalar field.

\section{Discussion}
In this construction, the test functions that are used to prepare a vector state of the quantum field partly define the interacting dynamics of the quantum field.
Although this structure may seem novel, infrared regularization for massless quantum fields is informally comparable to the formal approach taken here, in that the longest wavelengths that are measured are taken to determine the infrared dynamics, just as here properties of test functions that model an experimental apparatus determine the dynamics at all frequencies.
Also comparably, regularization by momentum cutoff, for example, generally takes the cutoff scale to be determined by the highest frequencies of the test functions that are used to model an experimental apparatus, introducing a vaguely defined weak nonlinear dependency on the test functions, and all constructions of an effective dynamics introduce length scales that are informally determined by the experimental context.
Although the principal example here has taken the high frequency properties of the interaction dynamics to depend on the large scale properties of the experimental apparatus, the functional relationship between test functions and the interaction dynamics can be tuned in a wide variety of ways, describing a complex overall conditioning of space--time by an experimental apparatus (with the bringing together and running of an experimental apparatus generally taking months or years in contrast to the nanosecond--scale creations of individual measurement records, but the conditioning is presumably, for most experiments, rather weak).

Other regularizations, by comparison, are either less physically motivated than regularization by test function (for example, Pauli-Villars regularization introduces ``indefinite metric sectors''\cite[p320]{IZ}, which removes the possibility of an elementary probability interpretation for the regularized theory, and dimensional regularization as it is usually presented is an essentially mathematical procedure), are not Poincar\'e covariant constructions (for example, lattice and momentum cutoff regularizations), or do not preserve microcausality (for example, point--splitting regularizations).

There are very many possibilities when we introduce quantum field operators as nonlinear functionals of test functions, even under the constraints of Poincar\'e and internal symmetries, so that we are once again in the position of classical physics, with far more models than we can fully investigate.
Although this is in many ways unwelcome, quantum field theory introduces functionals over Schwartz space into an algebra of operators, so it is to be expected that we will have to work with a wider variety of models than when we work with functions over Minkowski space in classical field theory, so that more than the Lagrangian that is sufficient to specify a dynamics of a classical field theory is required to specify a dynamics in a functional context.

This formalism provides a means for modeling systems using test functions that may be very different from plane waves.
It suggests, in particular, a systematic study of whether or how the dynamics of a quantum field can usefully be nonlinearly dependent on all aspects of the test functions used to model an experimental apparatus.
Indeed, determining the dependence of experimental results on changes to large--scale properties of apparatus is a project that is little suggested by conventional HEP, where low energy components are of relatively little importance.
It should be emphasized that the approach here is not intended as the only way to construct a model ---indeed it is much to be hoped that a different approach can restrict the many possibilities and be more tractable--- but only as a way of constructing mathematically more robust models that has not previously been pursued and that can approach the phenomenology of perturbative quantum field theory when states are prepared and measured as close to plane wave states.

\appendix
\section{A heuristic discussion of a conventional interacting quantum field}\label{IZAppendix}
\newcommand\iD{{\mathrm{i}\hspace{-1.5pt}\Delta}}
The direct textbook way to construct an interacting scalar quantum field is to introduce a formal time--dependent transformation of a free quantum field $\hat\phi(x)\;$~\cite[\S 6-1-1]{IZ},
\begin{eqnarray}\label{XiDefinition}
  \hat\xi(x)&=&\hat U^{-1}(x_0)\hat\phi(x)\hat U(x_0)\hspace{6em}\mathrm{where}\ 
      \hat U(\tau)=\mathsf{T}\!\left[\!\!\raisebox{-0.9ex}{
                       $\rme^{-\rmi\!\!\!\int\limits_{-\infty}^\tau\!\!\!\hat H(y)\Intd^4y}$}\right],\cr
            &=&\mathsf{T}^\dagger\!\left[\!\!\raisebox{-0.9ex}{
                       $\rme^{-\rmi\!\!\!\int\limits_{-\infty}^\infty\!\!\!\hat H(y)\Intd^4y}$}\right]
               \mathsf{T}\!\left[\!\!\raisebox{-0.9ex}{
                       $\hat\phi(x)\rme^{-\rmi\!\!\!\int\limits_{-\infty}^\infty\!\!\!\hat H(y)\Intd^4y}$}\right]
    =\mathsf{T}^\dagger\!\left[\rme^{-\rmi\hat\mathcal{L}}\right]
          \mathsf{T}\!\left[\hat\phi(x)\rme^{-\rmi\hat\mathcal{L}}\right],
\end{eqnarray}
where $\hat H(y)$ is a positive semi--definite local interaction operator, constructed as a sum of normal--ordered products of $\hat\phi(y)$ and its derivatives.
It is not generally noted that there are components of $\hat U(x_0)$ that are space--like separated from $x$, which therefore commute with $\hat\phi(x)$, and, because of time--ordering, those components cancel with the time-reversed components of the inverse
$\hat U^{-1}(x_0)$.
Consequently, the interacting field $\hat\xi(x)$ can also be written Lorentz invariantly as
\begin{equation}\label{xiDefinition}
  \hat\xi(x)=\mathsf{T}^\dagger\!\left[\rme^{-\rmi\hat\mathcal{L}(x)}\right]\hat\phi(x)
                          \mathsf{T}\!\left[\rme^{-\rmi\hat\mathcal{L}(x)}\right],\qquad
  \mathrm{where}\ \hat\mathcal{L}(x)= \int\limits_{\blacktriangle(x)}\!\hat H(y)\Intd^4y
\end{equation}
and $\blacktriangle(x)=\{y:(x-y)^2\ge 0\ \mathrm{and}\ x_0>y_0\}$ is the region of space-time that is light--like or time--like separated from and earlier than $x$.

It is of interest both to see the relationship with the classical differential equation that is related to the interaction Lagrangian and to make explicit that the operator $\hat\xi(x)$ is a complex of operators associated with the backward light--cone.
The adjoint action of $\hat\phi(x)$ on a time--ordered expression is a derivation, because time--ordering ensures commutativity, so that, taking a quartic scalar interaction with Hamiltonian density $\SmallFrac{\lambda}{4!}\!:\!\hat\phi^4(y)\!:$ as an example,
$$
  \left[\hat\phi(x),\mathsf{T}\!\left[\left(\int\!:\!\hat\phi^4(y)\!:\Intd^4y\right)^n\right]\right]=
       \mathsf{T}\!\left[\int\!4n\iD(x-z):\!\hat\phi^3(z)\!:\Intd^4z\left(\int\!:\!\hat\phi^4(y)\!:\Intd^4y\right)^{n-1}\right],
$$
where
\begin{equation}\label{iDG}
  \iD(x-z)=-\rmi(G_{\mathrm{ret}}(x-z)-G_{\mathrm{adv}}(x-z))=[\hat\phi(x),\hat\phi(z)]
\end{equation}
is the free field commutator and $G_{\mathrm{ret}}(x-z)$ and $G_{\mathrm{adv}}(x-z)$ are the retarded and advanced Green functions~\cite[\S 1-3-1]{IZ}.
For the interacting field, therefore, we have the construction
\begin{eqnarray}
  \hat\xi(x)&=&\mathsf{T}^\dagger\!\left[\rme^{-\rmi\hat\mathcal{L}(x)}\right]\hat\phi(x)
                          \mathsf{T}\!\left[\rme^{-\rmi\hat\mathcal{L}(x)}\right]\cr
            &&\hspace{3em}\mbox{[$\hat\phi(x)$ acts as a derivation, ...]}\cr
            &=&\mathsf{T}^\dagger\!\left[\rme^{-\rmi\hat\mathcal{L}(x)}\right]
                       \left(\!\mathsf{T}\left[\rme^{-\rmi\hat\mathcal{L}(x)}\right]\hat\phi(x)
                          -\mathsf{T}\!\left[\SmallFrac{\mathrm{i\lambda}}{3!}\!\!\!
                              \int\limits_{\blacktriangle(x)}\!\iD(x-z):\!\hat\phi^3(z)\!:\Intd^4z
                                    \rme^{-\rmi\hat\mathcal{L}(x)}\right]\right)\cr
            &=&\hat\phi(x)-\mathsf{T}^\dagger\!\left[\rme^{-\rmi\hat\mathcal{L}(x)}\right]
                          \mathsf{T}\!\left[\SmallFrac{\mathrm{i\lambda}}{3!}\!\!\!
                              \int\limits_{\blacktriangle(x)}\!\iD(x-z):\!\hat\phi^3(z)\!:\Intd^4z
                                    \rme^{-\rmi\hat\mathcal{L}(x)}\right]\cr
            &=&\hat\phi(x)-\SmallFrac{\mathrm{i\lambda}}{3!}\!\!\!\int\limits_{\blacktriangle(x)}\iD(x-z)
                                \mathsf{T}^\dagger\!\left[\rme^{-\rmi\hat\mathcal{L}(x)}\right]
                                \mathsf{T}\!\left[:\!\hat\phi^3(z)\!:\rme^{-\rmi\hat\mathcal{L}(x)}\right]\Intd^4z\cr
            &&\hspace{3em}\mbox{[components of $\hat\mathcal{L}(x)$ that are space-like separated from $z$}\cr
            &&\hspace{13em}\mbox{or are later than $z$ \emph{cancel}, leaving $\hat\mathcal{L}(z)$, ...]}\cr
            &=&\hat\phi(x)-\SmallFrac{\mathrm{i\lambda}}{3!}\!\!\!\int\limits_{\blacktriangle(x)}\iD(x-z)
                                \mathsf{T}^\dagger\!\left[\rme^{-\rmi\hat\mathcal{L}(z)}\right]
                                     :\!\hat\phi^3(z)\!:
                                \mathsf{T}\!\left[\rme^{-\rmi\hat\mathcal{L}(z)}\right]\Intd^4z\cr
            &=&\hat\phi(x)-\SmallFrac{\mathrm{i\lambda}}{3!}\!\!\!
                              \int\limits_{\blacktriangle(x)}\!\iD(x-z):\!\hat\xi^3(z)\!:\Intd^4z\quad
        \biggl[:\!\hat\xi^3(z)\!: \eqdef\mathsf{T}^\dagger\!\left[\rme^{-\rmi\hat\mathcal{L}(z)}\right]
                                     :\!\hat\phi^3(z)\!:
                                \mathsf{T}\!\left[\rme^{-\rmi\hat\mathcal{L}(z)}\right]\biggr]\cr
            &=&\hat\phi(x)-\SmallFrac{\lambda}{3!}\!\!
                              \int\!G_{\mathrm{ret}}(x-z):\!\hat\xi^3(z)\!:\Intd^4z,\label{IntegralEqn}
\end{eqnarray}
where the restriction to the backward light--cone $\blacktriangle(x)$ is equivalent to replacing the propagator $\iD(x-z)$ by $-\rmi G_{\mathrm{ret}}(x-z)$, as we see from Eq. (\ref{iDG}).
Insofar as we can take $:\!\hat\phi^3(z)\!:$ formally to be an infinite multiple of $\hat\phi(z)$ subtracted from $\hat\phi^3(z)$, we can take $:\!\hat\xi^3(z)\!:$ formally to be the same infinite multiple of $\hat\xi(z)$ subtracted from $\hat\xi^3(z)$.
$\hat\phi(x)$ satisfies the homogeneous Klein-Gordon equation, $(\Box+m^2)\hat\phi(x)=0$, and $G_{\mathrm{ret}}(x-z)$ satisfies $(\Box+m^2)G_{\mathrm{ret}}(x-z)=\delta^4(x-z)$, so, applying the operator $(\Box+m^2)$ to the last line of Eq. (\ref{IntegralEqn}), $\hat\xi(x)$ satisfies the nonlinear differential equation
\begin{eqnarray}\label{FieldDiffEqn}
  (\Box+m^2)\hat\xi(x)+\SmallFrac{\lambda}{3!}:\!\hat\xi^3(x)\!:\;=0.
\end{eqnarray}

The above construction is very heuristic, but it shows that apart from the mathematical necessity to regularize we can construct an interacting field by replacing $\hat\phi(x)$ at a point by a complex of operators at points of the backward light--cone of $x$, constructed using the propagator $G_{\mathrm{ret}}(x-z)$.
We have also shown that if we approach an interacting quantum field theory in this way we can take virtual particles, the free particle propagator modified by the effect of time ordering, effectively to be associated with the retarded Greens function instead of with the Feynman propagator.


\begin{thebibliography}{9}
\bibitem{Haag}
  {R. Haag, \textit{Local Quantum Physics}, Springer, Berlin (1996).}
\bibitem{MorganHadamard}
  {P. Morgan, arXiv:1211.2831v2 [math-ph].}
\bibitem{IZ}
  {C. Itzykson and J.-B. Zuber, \textit{Quantum Field Theory}, McGraw-Hill International Edition, Singapore (1985).}
\end{thebibliography}
\end{document}